\documentclass[11pt,twoside]{article}
\usepackage{asp2004}
\usepackage{psfig}
\usepackage{epsfig}
\usepackage{aas_macros}
\begin{document}

\title{What can we learn about black-hole
  formation from black-hole X-ray binaries?}
\author{Gijs Nelemans}
\affil{Institute of Astronomy, University of Cambridge, UK\\
and\\
Department of Astrophysics, Radboud University Nijmegen,  NL}

\begin{abstract}
I discuss the effect of the formation of a black hole on a (close)
binary and show some of the current constraints that the observed
properties of black hole X-ray binaries put on the formation of black
holes. In particular I discuss the evidence for and against asymmetric
kicks imparted on the black hole at formation and find contradicting
answers, as there seems to be evidence for kick for individual systems
and from the Galactic $z$-distribution of black hole X-ray binaries,
but not from their line-of-sight velocities.
\end{abstract}
\thispagestyle{plain}

\section{Introduction: What do we want to learn?}

We can study black holes only when they interact with other stars or
material. For stellar mass black holes the only way to study them is
when they reside in binaries and in particular when they accrete
material from their (low-mass) companion stars in black-hole X-ray
binaries (BHXBs). These binaries can be used to get information about
the formation of black holes, of course with the caveat that strictly
speaking they only constrain the formation of black holes \textit{in
binaries}. The formation of black holes from massive stars is poorly
understood \citep[e.g.][]{fk01}. The questions we would like to answer
are:\\ i) Which stars form black holes? (What is the minimum mass of a
main sequence star that forms a black hole, and does this depend on
binarity/rotation etc?)\\ ii) Do black holes form in either direct
collapse, a supernova explosion or either?  \\ iii) How much mass is
lost during the black hole formation?  (which is important for the
chemical enrichment of the interstellar medium) \\ iv) Is the black
hole formation symmetric or asymmetric?\\ Here I will discuss only the
last three questions.

\section{Effects of the formation of a black hole on the binary}

\begin{figure}[t]
\begin{center}
\psfig{figure=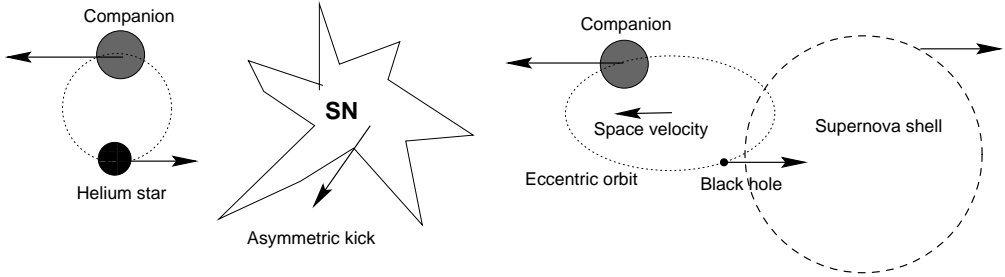,width=\textwidth}
\end{center}
\vspace*{-0.5cm}
\caption{Effects of black hole formation on the binary (see text).}
\label{fig:binary}
\vspace*{-0.3cm}
\end{figure}

I discuss four ways in which the formation of a black hole in a binary
can influence the companion and the binary as a whole (see
Fig.~\ref{fig:binary})\\
\textbf{Pollution companion}\\ If the formation of the black hole is
  accompanied by explosive mass loss, part of this can be captured by the
  companion and, if enough is retained in layers near the surface, can
  cause the companion to show peculiar abundances
  \citep{irb+99,pnm+02,gri+04}\newpage
\noindent \textbf{Space velocity}\\ If mass is lost quickly it will
  carry linear momentum, causing the centre of mass of the remaining
  binary to move in the opposite direction. Furthermore, if the
  formation of the black hole is asymmetric a kick will be imparted on
  the black hole, which also causes a space velocity ($v_{\rm sys}$)
  of the binary.
\\ \textbf{Eccentric orbit}\\ Both explosive mass loss and a
  kick induce an eccentricity ($e$)in the binary.
\\ \textbf{Angle  between $v_{\rm sys}$ and orbit}\\ If the kick on the black hole is
  directed out of the orbital plane, the direction of the resulting
  space velocity of the binary will make an angle with the orbital
  plane. Symmetric mass loss always results in a space velocity directed in the
  plane of the orbit.

\subsection{Symmetric mass loss}

In the case of symmetric mass loss, so in absence of a kick, there is
a unique relation betweem $v_{\rm sys}$ and $e$, as both are uniquely
determined by the amount of mass loss\footnote{Be X-ray binaries, in
which a neutron star accretes (periodically) from a Be star do not
follow this relation, which is direct evidence that neutron stars do
receive kicks at formation \citep{vpb+00}. Interestingly, there are no
black-hole Be-X-ray binaries...}. If after the formation of the black
hole the orbit of the binary is small enough that the companion will
start mass transfer to the black hole (and thus we can observe the
binary), the eccentric orbit quickly circularizes
\citep[e.g.][]{kal99}, yielding the following relation between the
space velocity of the binary and the observable parameters
\citep[cf.][]{nth99}
\begin{displaymath}
v_{\rm sys}
 = 213 
 \! \left(\!\frac{\Delta M}{M_{\odot}}\!\right) \!\!\left(\!\frac{m}{M_{\odot}} 
\!\right) \!\!\left(\! \frac{P_{\rm re-circ}}{\mbox{day}} \!\right)^{\!\!\!-\frac{1}{3}} \!\!\left(
\frac{\!M_{\rm BH}\!+\!m}{M_{\odot}} \!\right)^{\!\!\!-\frac{5}{3}}\! \mbox{km s$^{-1}$}
\end{displaymath}

\begin{figure}[t]
\begin{center}
\psfig{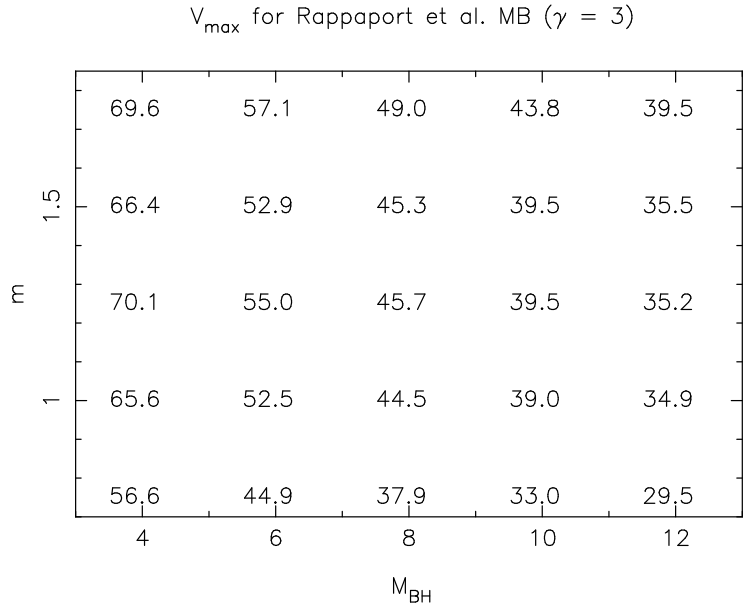}
\end{center}
\vspace*{-0.5cm}
\caption{Maximum space velocities of LMBBs as function of their masses
  at the onset of mass transfer. From \citet{nvb04}}
\label{fig:Vmax}
\vspace*{-0.3cm}
\end{figure}

Based on the properties of binaries that can reach this phase of mass
transfer, a maximum space velocity can be derived for short period
black hole binaries with low-mass companions (LMBBs), see
\citet{nvb04}. The maximum velocities as a function of the system
parameters \textit{at the onset of mass transfer} (which generally is
very different from the current parameters) is shown in
Fig.~\ref{fig:Vmax}.

\section{Galactic $z$-distribution}

For most systems it is not possible to determine the space velocity
and the system parameters, let alone infer what they were at the
onset of mass transfer. However, \citet{wp96} noticed that the average
$z$-height of BHXBs was about half that of neutron star X-ray binaries
(NSXBs) suggesting a lower average velocity of the BHXBs. They
therefore concluded that there was no evidence for asymmetric kicks
imparted on black holes, in contrast with neutron stars.

\begin{figure}[t]
\begin{center}
\psfig{figure=nelemans_f3.ps,width=0.49\textwidth}
\psfig{figure=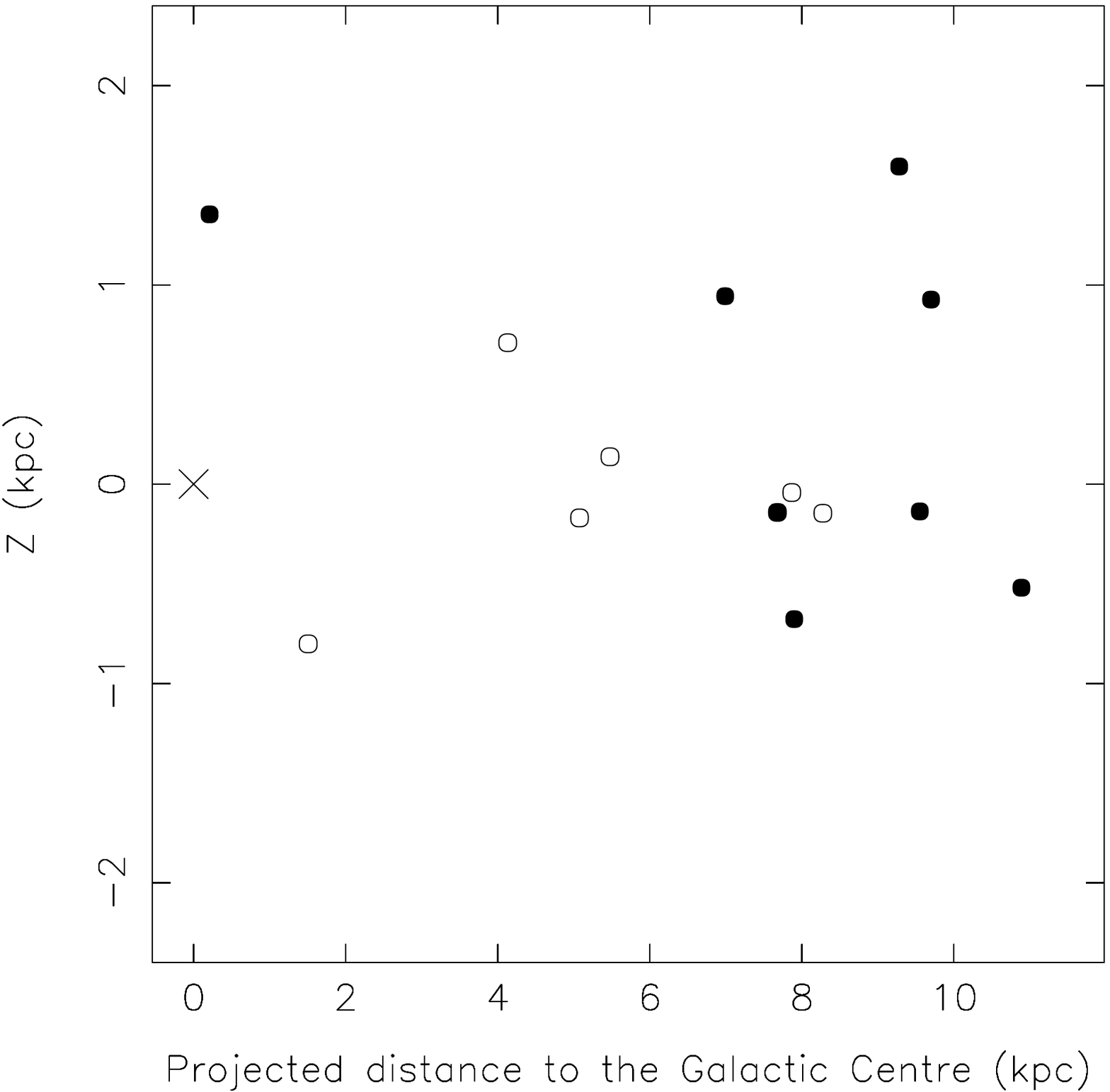,width=0.49\textwidth}\\
\psfig{figure=nelemans_f5.ps,width=0.35\textwidth,angle=-90,clip=}
\psfig{figure=nelemans_f6.ps,width=0.35\textwidth,angle=-90,clip=}\\
\end{center}
\vspace*{-0.5cm}
\caption{Top left: distances from the Galactic centre and the Galactic
  plane of BHXBs (solid circles) and NSXBs (open symbols), from
  \citet{jn04}. Top right: only LMBBs. Bottom: comparison between
  expected $R-z$ distribution of LMBBs with velocities according to
  Fig.~\ref{fig:Vmax} (left) and with kick of 110 km/s (right).}
\label{fig:distGC}
\vspace*{-0.3cm}
\end{figure}

We recently repeated this analysis, using new distances and newly
discovered systems and found that the current situation is that there
is no difference anymore $z_{\rm rms, BH} = z_{\rm rms, NS} \approx 1$
kpc \citep[see Fig.~\ref{fig:distGC}]{jn04}. This in principle
suggests that black holes do get kicks at birth, but there are a
number of selections effects that complicate the comparison.  Firstly,
the black hole systems are relatively close by (as one needs to see
the companion in order to measure its radial velocity and thus
determine the mass of the compact object). This means the BHXBs are
farther from the Galactic centre and thus move in a lower
potential. Secondly, there are differences in formation between NSXBs
and BHXBs, that can give rise to quite different velocities, even if
the mass loss was symmetric.

As a simple experiment, I plot in Fig.~\ref{fig:distGC} (top right)
the $R-z$ distribution of the subset of systems that are LMBBs (solid
circles), and in the bottom two panels their expected distributions in
case of symmetric mass loss, i.e. with velocities as in
Fig.~\ref{fig:Vmax} (bottom left panel) and in case they receive a
kick of 110 km/s (bottom right panel). Except for the system close to
the Galactic centre, all seem to be consistent with symmetric mass
loss.


\section{Observations of 3D velocities}

\begin{figure}[t]
\begin{center}
\psfig{figure=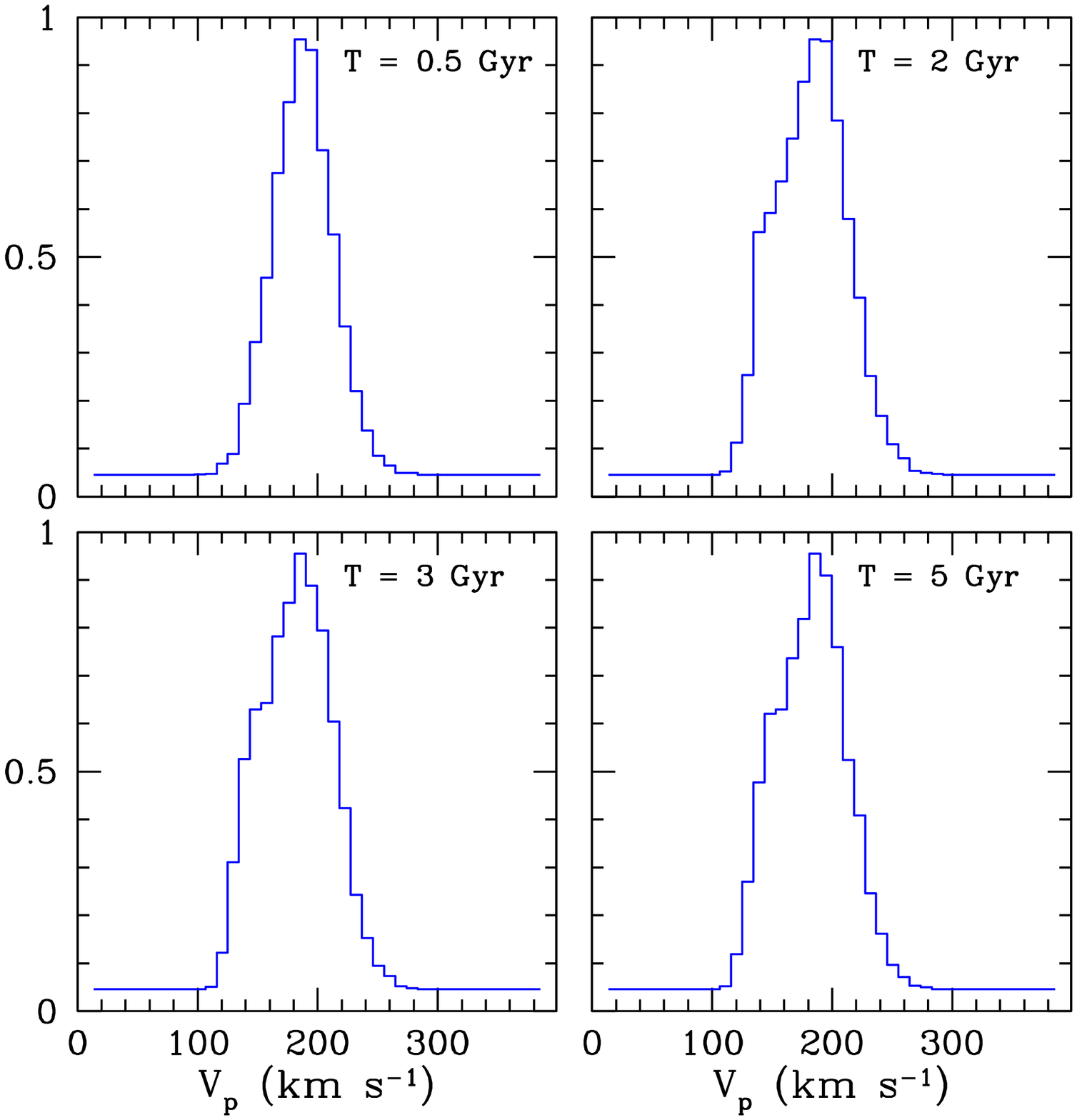,width=0.45\textwidth,clip=}
\psfig{figure=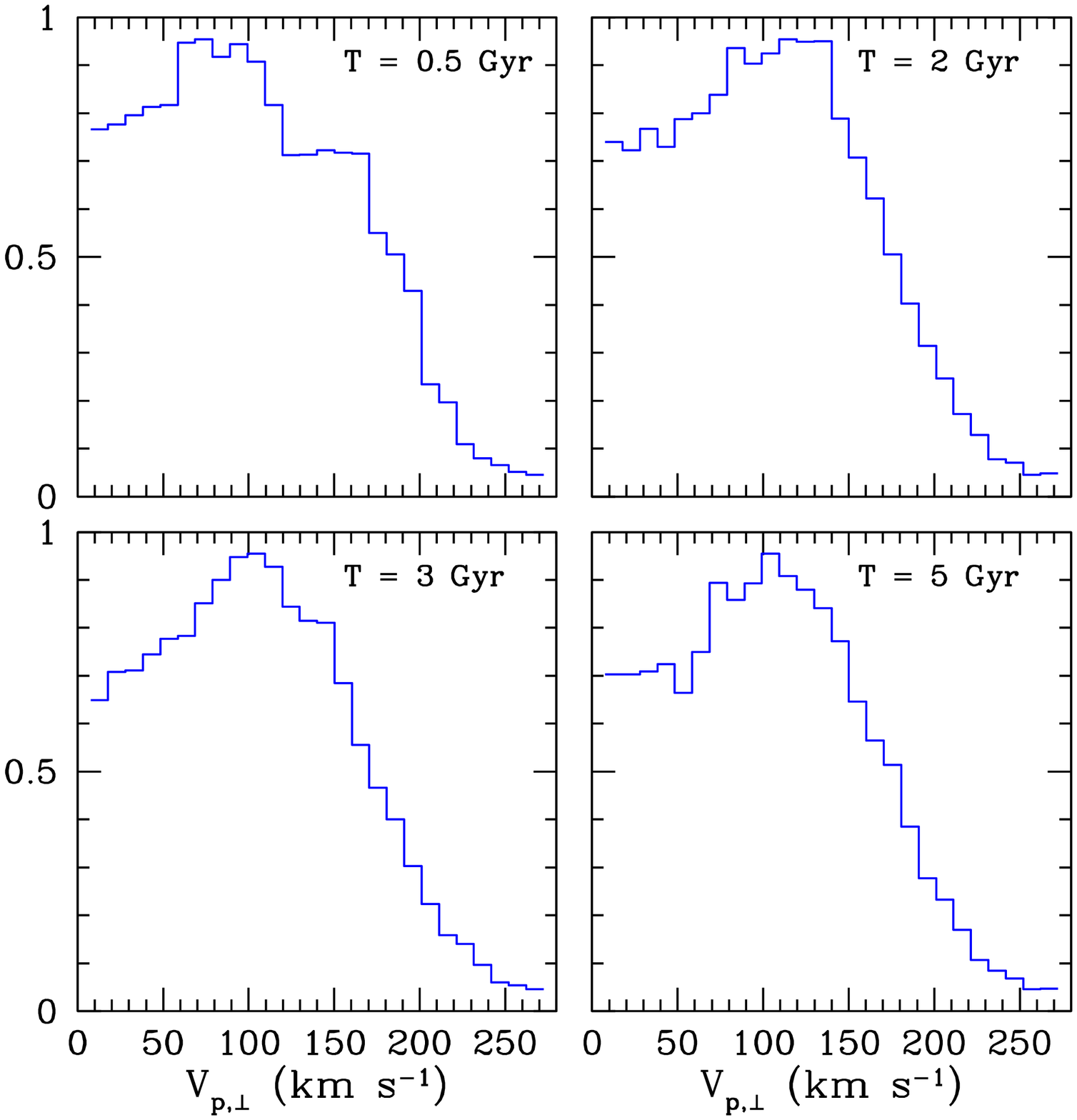,width=0.45\textwidth,clip=}\\
\end{center}
\vspace*{-0.5cm}
\caption{Peculiar velocity of XTE J1118+480 after the formation of the
  black hole, and its component perpendicular to the orbital plan
  based on the current system parameters \citep[from][]{gcp+04}. }
\label{fig:alessia_vpec}
\vspace*{-0.3cm}
\end{figure}

For most systems we know their period, system radial velocity, their
masses and the binary inclination. In addition sometimes a radio or
optical proper motion is know \citep{mdm+01,mmr+02,mr03}. In principle
this can be used to get a 3D velocity by combining the proper motion
with the (generally very uncertain) distance to get the transverse
velocity and the radial velocity plus local standard of rest (which
again depends on the distance) to get the peculiar radial velocity. In
order to constrain the black hole formation one also needs to estimate
the age of the system, in order to trace the orbit back in the
Galactic potential to find the system parameters after the formation
of the black hole. A number of systems have recently been analyzed in
this way\\ \textbf{XTE J1118+480}\\ \citet{gcp+04} studied the
formation of XTE J1118+480 and traced back its orbit. They show that
the memory of its initial position is quickly lost. However,
Fig.~\ref{fig:alessia_vpec} shows that even after 5 Gyr, the peculiar
velocity of the binary after the formation of the black hole is still
quite well constrained and is significantly above the maximum velocity
that can be reached with symmetric mass loss for any of the system
parameters leading to LMBBs (cf. Fig.~\ref{fig:Vmax}). Also the
component of the peculiar velocity perpendicular to the orbital plane
is likely non-zero (Fig.~\ref{fig:alessia_vpec}). Both findings
suggest the black hole received a kick.

\noindent \textbf{GRO J1655-40}\\ Kalogera \& Willems (in prep.)
  performed a similar analysis of the intermediate mass X-ray binary
  GRO J1655-40 and found solutions without kick, but only when
  allowing the binary to get circularized very quickly after the
  formation of the black hole (for details see Kalogera \& Willems, in
  prep.).

\section{Radial velocities}

Finally, if kicks of the order of 100 km/s or higher indeed are
commonly imparted on black holes, one would expect the current
peculiar line-of-sight velocities (i.e. compared to their local
standard of rest) to be at least several tens of km/s. In Table~1 I
show the line-of-sight velocities of BHXBs, together with rough values
of the system parameters \citep[see e.g.][for recent
reviews]{oro03,cc04,mr04}. These velocities are remarkably low and to
me suggest that certainly not all BHXBs received (large) kicks.

\begin{table}[t]
\begin{center}
{\small
\begin{tabular}{@{}llrlrr}
Name    &$P$     &M$_{\rm BH}$    &M$_{\rm comp}$   &$z$    &$v_{\rm
  sys, los}$ \\         
 & (hr) & & & pc & km/s \\ \hline \hline
XTE J1118+480    & 4.08   &7     &0.25     &1600  &$-$7.2\\         
GRO J0422+32    & 5.09   &4     &0.3      &$-$525  &24.2  \\ 
GS 1009-45    & 6.84   &5     &0.7      &925   &10.8   \\
A0620-00    & 7.75   &11    &0.6      &$-$125  &$-$9.5   \\
GS 2000+25    & 8.27   &8     &0.5      &$-$150  &$-$3.5   \\
XTE J1859+226    & 8.61   &8     &        &950   &      \\
GS 1124-683    & 10.4   &7     &0.8      &$-$675  &36.3   \\
H1705-250    & 12.5   &7     &0.3      &1350  &166.7  \\ \hline
4U 1543-47    & 26.8   &9     &2.5      &700   &17.7   \\
XTE J1550-564    & 37     &10    &$<$1     &$-$175  &12.9   \\
GX339-4     & 42.1   &$>$2    &      &$-$450  &      \\
GRO J1655-40    & 62.9   &6     &2.8      &125   &$-$116.8\\ 
SAX J1819.3-2525    & 67.6   &7     &3.1      &$-$800  &$-$2.2   \\
GS 2023+338    & 155.3  &12    &0.6      &$-$150  &$-$6.2   \\
GRS 1915+105    & 816    &14    &1.2      &$-$50   &$-$16.6  \\ \hline
Cyg X-1     & 134.4  &8     &15       &100   &$-$12.6
\end{tabular}}
\end{center}
\caption{Rough system parameters and observed velocities of BHXBs. Top rows are LMBBs, middle have higher companion masses and/or longer periods, while Cyg X-2 has a high-mass companion.}
\end{table}

\section{Conclusions}

By analyzing the properties of BHXBs a number of things can be learned
about the formation of black holes. The first is that some BHs are
formed in a supernova, with (quite some) mass loss. Secondly, the
$z$-distribution BHXBs and NSXBs are similar, which naively suggests
kicks are imparted on black holes as well as on neutron stars but this
needs detailed investigation.  The measurement of 3D velocities opens
the possibility to study BH formation in \textit{detail} and the first
results indicate that black holes were formed in a supernova and
received an asymmetric kick. However, line-of-sight velocities of
BHXBs are generally low, suggesting not all black holes receive a
kick.


\begin{thebibliography}

\bibitem[{Charles \& Coe(2004)}]{cc04}
Charles P., Coe M., 2004, in Lewin W., van~der Klis M., eds., Compact Stellar
  X-Ray Sources, CUP

\bibitem[{{Fryer} \& {Kalogera}(2001)}]{fk01}
{Fryer} C.L., {Kalogera} V., 2001, \apj,  554, 548

\bibitem[{{Gonz{\' a}lez Hern{\' a}ndez} et~al.(2004){Gonz{\' a}lez Hern{\'
  a}ndez}, {Rebolo}, {Israelian} et~al.}]{gri+04}
{Gonz{\' a}lez Hern{\' a}ndez} J.I., {Rebolo} R., {Israelian} G., et~al., 2004,
  \apj,  609, 988

\bibitem[{{Gualandris} et~al.(2004){Gualandris}, {Colpi}, {Zwart} \&
  {Possenti}}]{gcp+04}
{Gualandris} A., {Colpi} M., {Zwart} S.P., {Possenti} A., 2004, \apj, in
  press, astro-ph/0407502

\bibitem[{Israelian et~al.(1999)Israelian, Rebolo, Basri, Casares \&
  Mart\'{\i}n}]{irb+99}
Israelian G., Rebolo R., Basri G., Casares J., Mart\'{\i}n E.L., 1999, Nat,
  401, 142

\bibitem[{{Jonker} \& {Nelemans}(2004)}]{jn04}
{Jonker} P.G., {Nelemans} G., 2004, \mnras, in press

\bibitem[{Kalogera(1999)}]{kal99}
Kalogera V., 1999, ApJ,  521, 723

\bibitem[{McClintock \& Remillard(2004)}]{mr04}
McClintock J., Remillard R.A., 2004, in Lewin W., van~der Klis M., eds.,
  Compact Stellar X-Ray Sources, CUP

\bibitem[{{Mirabel} \& {Rodrigues}(2003)}]{mr03}
{Mirabel} I.F., {Rodrigues} I., 2003, Science,  300, 1119

\bibitem[{{Mirabel} et~al.(2001){Mirabel}, {Dhawan}, {Mignani}, {Rodrigues} \&
  {Guglielmetti}}]{mdm+01}
{Mirabel} I.F., et al.,
  2001, \nat,  413, 139

\bibitem[{{Mirabel} et~al.(2002){Mirabel}, {Mignani}, {Rodrigues}
  et~al.}]{mmr+02}
{Mirabel} I.F., {Mignani} R., {Rodrigues} I., et~al., 2002, \aap,  395, 595

\bibitem[{Nelemans et~al.(1999)Nelemans, Tauris \& van~den Heuvel}]{nth99}
Nelemans G., Tauris T.M., van~den Heuvel E.P.J., 1999, A\&A,  352, L87

\bibitem[{Nelemans et~al.(2004)Nelemans, van~den Heuvel \&
  Bhattacharya}]{nvb04}
Nelemans G., van~den Heuvel E.P.J., Bhattacharya D., 2004, in van~den Heuvel
  , Kaper, Rol, Wijers, eds., \emph{ASP conf. ser.} 308, ASP, San
  Francisco, p. 155

\bibitem[{{Orosz}(2003)}]{oro03}
{Orosz} J.A., 2003, in IAU Symposium 212, p. 365

\bibitem[{{Podsiadlowski} et~al.(2002){Podsiadlowski}, {Nomoto}, {Maeda}
  et~al.}]{pnm+02}
{Podsiadlowski} P., {Nomoto} K., {Maeda} K., et~al., 2002, \apj,  567, 491

\bibitem[{{van den Heuvel} et~al.(2000){van den Heuvel}, {Portegies Zwart},
  {Bhattacharya} \& {Kaper}}]{vpb+00}
{van den Heuvel} E.P.J., {Portegies Zwart} S.F., {Bhattacharya} D., {Kaper} L.,
  2000, \aap,  364, 563

\bibitem[{White \& van Paradijs(1996)}]{wp96}
White N.E., van Paradijs J., 1996, ApJ,  473, L25

\end{thebibliography}

\end{document}